\documentclass[12pt]{article}
\pdfoutput=1 

\usepackage{jheppub} 
\usepackage{graphicx}  
\usepackage{dcolumn}   
\usepackage{bm}        
\usepackage{amssymb}   
\usepackage{subfigure}
\usepackage{epstopdf}
\usepackage[dvipsnames]{xcolor}
\usepackage{slashed}
\usepackage[utf8]{inputenc}
\usepackage{multirow}
\usepackage{footnote}
\usepackage{amsmath}
\allowdisplaybreaks[4]
\usepackage{accents}
\newlength{\dhatheight}

\usepackage{hyperref}

\def\figureautorefname~#1\null{Fig.\,#1\null}
\def\tableautorefname~#1\null{Tab.\,#1\null}

\def\equationautorefname~#1\null{Eq.\,(#1)\null}

\title{Tree-level Interference in VBF production of Vh}
\author{Daniel Stolarski and}
\author{Yongcheng Wu}
\affiliation{Ottawa-Carleton Institute for Physics, Carleton University, 1125 Colonel By Drive, Ottawa, Ontario K1S 5B6, Canada}
\emailAdd{stolar@physics.carleton.ca}
\emailAdd{ycwu@physics.carleton.ca}

\abstract{
Vector boson scattering is a well known probe of electroweak symmetry breaking. Here we study a related process of two electroweak vector bosons scattering into a vector boson and a Higgs boson ($VV \rightarrow Vh, V=W,Z$). This process exhibits tree level interference and grows with energy if the Higgs couplings to electroweak bosons deviate from their Standard Model values. Therefore, this process is particularly sensitive to the relative sign of the ratio of the coupling between the Higgs and the $W$ and $Z$, $\lambda_{WZ}$. In this work we show that a high energy lepton collider is well suited to study this process through vector boson fusion, estimate the potential sensitivity to this ratio, and show that a relatively modest amount of data can exclude 
$\lambda_{WZ} \simeq -1$. 
}

\begin{document}
\titlepage
\maketitle
\newpage

\flushbottom

\section{Introduction}

In the Standard Model (SM), the Higgs boson is a necessary ingredient to keep longitudinal gauge boson scattering unitary at high energy~\cite{LlewellynSmith:1973yud,Veltman:1976rt,Lee:1977yc,Lee:1977eg,Passarino:1985ax,Passarino:1990hk}. Even in the presence of the Higgs, if its couplings deviate from those predicted by the SM, longitudinal gauge boson scattering will still grow with energy and new physics is needed to make the theory unitary~\cite{Cornwall:1974km,Giudice:2007fh,Contino:2010mh}. Therefore, the experimental discovery of the Higgs~\cite{Aad:2012tfa,Chatrchyan:2012xdj} is insufficient to probe the nature of the unitiarization of longitudinal gauge boson scattering; detailed measurements of its couplings are necessary.

Vector boson scattering ($VV \rightarrow VV$, $V=Z,W$) is an extremely well studied subject~(for a review see \cite{Szleper:2014xxa}). One can do a rotation in gauge space to study the similar process $VV \rightarrow hh$~\cite{Contino:2010mh}, where $h$ is the Higgs boson. More complicated processes involving gauge bosons can also be used to measure various Higgs couplings even when the Higgs is not one of the external particles~\cite{Henning:2018kys}. In this work, we study another gauge rotated process: 
\begin{equation}
VV \rightarrow Vh, \; \; V=W,Z.
\label{eq:hardProcess}
\end{equation}
Like the two processes above, this one also exhibits growth with energy if the couplings of the Higgs to massive gauge bosons deviate from the Standard Model prediction. This process is especially sensitive to the ratio of the coupling of the Higgs to the $W$ relative to that of the $Z$. In particular, if we define $\kappa_W$ ($\kappa_Z$) as the deviation of the $W$ ($Z$) coupling to the Higgs from the SM prediction ($\kappa_W = \kappa_Z = 1$ in the SM, see~\autoref{eq:lag} below for more details), then we can define:
\begin{equation}
\lambda_{WZ} = \frac{\kappa_W}{\kappa_Z},
\label{eq:lambda}
\end{equation}
as the specified ratio. The process in~\autoref{eq:hardProcess} exhibits \textit{tree-level} interference effects between $W$ and $Z$ mediated processes, and the matrix element has a term that grows with energy proportional to $\lambda_{WZ} - 1$.

Because a heavy gauge boson collider is not feasible, the typical gauge boson scattering is studied experimentally via vector boson fusion (VBF), where $W$ or $Z$'s are radiated off the initial state fermions and then scatter off one another. Analytic understanding can be further gained via the Effective $W$ approximation~\cite{Dawson:1984gx,Kunszt:1987tk,Borel:2012by}, by treating the radiated gauge bosons as approximately on-shell. Gauge boson scattering has been observed at the LHC~\cite{Sirunyan:2017fvv,Sirunyan:2017ret,Aaboud:2018ddq,Aaboud:2019nmv}, but a lepton collider is a cleaner environment which may allow for more precise measurements. A lepton collider is a particularly good machine for precision studies of the Higgs, and planning has begun for several different machines including the ILC~\cite{Djouadi:2007ik,Baer:2013cma}, circular colliders CEPC~\cite{CEPCStudyGroup:2018ghi} and FCC-ee~\cite{Gomez-Ceballos:2013zzn,Abada:2019zxq}, and CLIC~\cite{Linssen:2012hp,Roloff:2018dqu} which has higher Center of Mass (CM) energies, with designs ranging from 1.5 to 3 TeV.\footnote{As this work was being completed, a study of this and other process at a high energy muon collider was posted in~\cite{Costantini:2020stv}.} 

Therefore, in this work we study the process of~\autoref{eq:hardProcess} with VBF at a high energy lepton collider:
\begin{align}
e^+\ e^- &\to \nu_e\ \bar{\nu}_e\ Z\ h, \nonumber\\
e^+\ e^- &\to \nu_e\ e\ W\ h.
\label{eq:VBFprocess}
\end{align}
This process grows with the center of mass energy of the lepton collider, with CLIC being a particularly good machine for its study. As we will show, this process is very sensitive to modification of $\lambda_{WZ}$ from the SM prediction because of the tree-level interference.  The work here is complimentary to others that use interference effects to measure $\lambda_{WZ}$ such as $e^+\ e^- \to W^+W^-h$ \cite{Chiang:2018fqf} or interference of tree and loop effects in $h\rightarrow 4\ell$~\cite{Chen:2016ofc}.

One particularly interesting scenario is that when $\lambda_{WZ}$ is negative relative to the SM prediction. Tree-level processes without interference effects such as decays of $h\rightarrow ZZ^*$~\cite{Sirunyan:2017exp,Aaboud:2017oem} and $h\rightarrow WW^*$~\cite{Sirunyan:2018egh,Aaboud:2018jqu} are only sensitive to $|\lambda_{WZ}|$. Fits to the couplings by the experimental collaborations~\cite{Khachatryan:2016vau,Sirunyan:2018koj,Aad:2019mbh} measure $\lambda_{WZ}$ with approximately 10\% precision but have almost no discriminating power between positive and negative values of $\lambda_{WZ}$.\footnote{The 13 TeV CMS analysis~\cite{Sirunyan:2018koj} actually has a best fit value that is negative, and
the 13 TeV ATLAS analysis~\cite{Aad:2019mbh} does not consider negative values of $\lambda_{WZ}$. }
The ultimate LHC sensitivity on this ratio is projected to be about 2\%~\cite{Cepeda:2019klc}, but as far as we are aware, there has been no study on the sensitivity to the sign from rate measurements at the LHC. Negative values of $\lambda_{WZ}$ can arise in models with scalars that have higher isospin representations~\cite{Low:2010jp} such as the Georgi-Machacek~\cite{Georgi:1985nv} model. In that case, however, the processes of~\autoref{eq:VBFprocess} will be many times larger than the SM prediction and can be easily measured with a high energy lepton collider. 

The remainder of this paper is structured as follows: in~\autoref{sec:2to2} we explore the behaviour of the $2 \rightarrow 2$ process focusing on the growth with energy, and in~\autoref{sec:VBF} we describe the process in vector boson fusion at a lepton collider and show the importance of tree-level interference. In~\autoref{sec:pheno} we conduct a phenomenological study including backgrounds and show how the analysis can be improved by taking differential distributions into account, and finally a summary is given in~\autoref{sec:sum}.

\section{2 \texorpdfstring{$\rightarrow$}{to} 2 Processes}
\label{sec:2to2}

We begin by analyzing the $2 \rightarrow 2$ processes
\begin{align}
W^+\ W^- &\rightarrow Z\ h, \nonumber\\
W^{\pm}\ Z\; &\rightarrow W^{\pm}\ h,
\label{eq:22process}
\end{align}
to understand the large interference effects that can occur. We parameterize the coupling of the Higgs ($h$) to the $W$ and $Z$ as
\begin{equation}
{\cal L}= h \left(  \kappa_W\, g\, m_W W_\mu^+W_\mu^- + \kappa_Z\, g\, \frac{m_Z^2}{2m_W} Z_\mu Z_\mu \right),
\label{eq:lag}
\end{equation}
where $g$ is the SM $SU(2)$ gauge coupling, and in the Standard Model $\kappa_W=\kappa_Z =1$, with values away from one parameterizing deviation from the SM prediction. 

Both processes in~\autoref{eq:22process} have contributions from $s$ and $t$-channel diagrams (the Feynman diagrams can be read from the VBF Feynman diagrams below in~\autoref{fig:WhZhFeynDia}). In the first (second) process, the $s$-channel diagram is proportional to $\kappa_Z$ ($\kappa_W$), while the $t$-channel diagrams are proportional to $\kappa_W$ ($\kappa_Z$). We can divide the process into sub-processes based on the number of transversely vs.~longitudinally polarized gauge bosons in this process. Direct calculation shows that the matrix elements with two or three longitudinal guage bosons grows with energy. This is analogous to the classic studies of $WW\rightarrow WW$ scattering~\cite{LlewellynSmith:1973yud,Veltman:1976rt,Lee:1977yc,Lee:1977eg,Passarino:1985ax,Passarino:1990hk}, and even though the $s$ and $t$-channel processes individually grow with energy, their sum displays a cancellation such that the final amplitude does not grow with energy and the theory remains unitary at high energy. We summarize the high energy behaviour of the polarized matrix elements in~\autoref{tab:HighEnXsec} for different combinations of vector polarizations. 

\begin{table}
\centering
\begin{tabular}{|c||c|c|c|c|}
\hline
 & $\mathcal{M}_{s/t}$  &  $d\sigma_{s/t}$ & $\mathcal{M}_{s}+\mathcal{M}_{t}$  &  $d\sigma_{\text{tot}}$ \\ \hline
TTT & $\frac{1}{\sqrt{s}}$ & $\frac{1}{s^2}$ & $\frac{1}{\sqrt{s}}$ & $\frac{1}{s^2}$\\ \hline
LTT & $s^0$ & $\frac{1}{s}$ & $s^0$ & $\frac{1}{s}$ \\ \hline
LLT & $\sqrt{s}$ & $s^0$ & $\frac{1}{\sqrt{s}}$ & $\frac{1}{s^2}$ \\ \hline
LLL & $s$ & $s$ & $s^0$ & $\frac{1}{s}$\\ \hline
\end{tabular}
\caption{High energy behaviour of the polarized process given in~\autoref{eq:22process} as a function of Mandelstam $s$ for the matrix element $\mathcal{M}$ and the differential cross section $d\sigma$. The first two columns are for the $s$- and $t$-channel processes individually, while the third and fourth columns are for the $s$- and $t$-channel processes summed. The different rows are for different combinations of polarizations of gauge bosons whether they are transverse (T) or longitudinal (L).  }
\label{tab:HighEnXsec}
\end{table}

We can analyze the case $W_LW_L\rightarrow Z_Lh$ in more detail. We expand the matrix elements for the $s$- and $t$-channel processes in the high energy limit:
\begin{align}
\mathcal{M}_s \left(W^+_L W^-_L \rightarrow Z_L h \right) =&  
\frac{\kappa_Z g^2 \cos\theta}{4  m_W^2}\left( s - m_h^2+2m_Z^2 \right) + \mathcal{O}\left(\frac{1}{s}\right), \\
\mathcal{M}_t \left(W^+_L W^-_L \rightarrow Z_L h \right) =& 
\frac{\kappa_Wg^2}{4 m_W^2}\left( \cos\theta\left(- s  + 2 m_W^2 + m_Z^2 - m_h^2 \right) + 8 m_W^2 \frac{\cos\theta}{\sin^2\theta} \right)  \nonumber\\
&+ \mathcal{O}\left(\frac{1}{s}\right),
\label{eq:22mat}
\end{align}
where $\theta$ is the scattering angle in the centre of momentum frame. The singularity in the forward ($\theta=0$) and backward ($\theta = \pi$) limits in the $t$-channel diagrams are artifacts of the high energy expansion and are cut off by masses in the full expression. Adding the two matrix elements:
\begin{equation}
\mathcal{M} \left(W^+_L W^-_L \rightarrow Z_L h \right) = \frac{g^2 \kappa_Z\cos\theta}{4 m_W^2}(1-\lambda_{WZ}) \, s +\mathcal{O}(s^0),
\label{eq:matLLL}
\end{equation}
where $\lambda_{WZ}$ is defined in~\autoref{eq:lambda} and equal to one in the SM. We see that at high energy, the SM predicts that the matrix element of this process goes to a constant, and thus cross section falls, preserving unitarity. On the other hand, if there is new physics that modifies the ratio of the coupling of the Higgs to the electroweak gauge bosons, $\lambda_{WZ}\neq 1$, then this amplitude and thus the cross section will grow quadratically with center of mass energy. This growth will eventually be cut off by new resonances or other effects of new physics. A particularly interesting case is that of $\lambda_{WZ} \simeq -1$. This is impossible to distinguish from the Standard model prediction without an interference measurement, and the process studied here is extremely sensitive to this scenario.

\begin{figure}[!tbp]
\centering
\includegraphics[width=.49\textwidth]{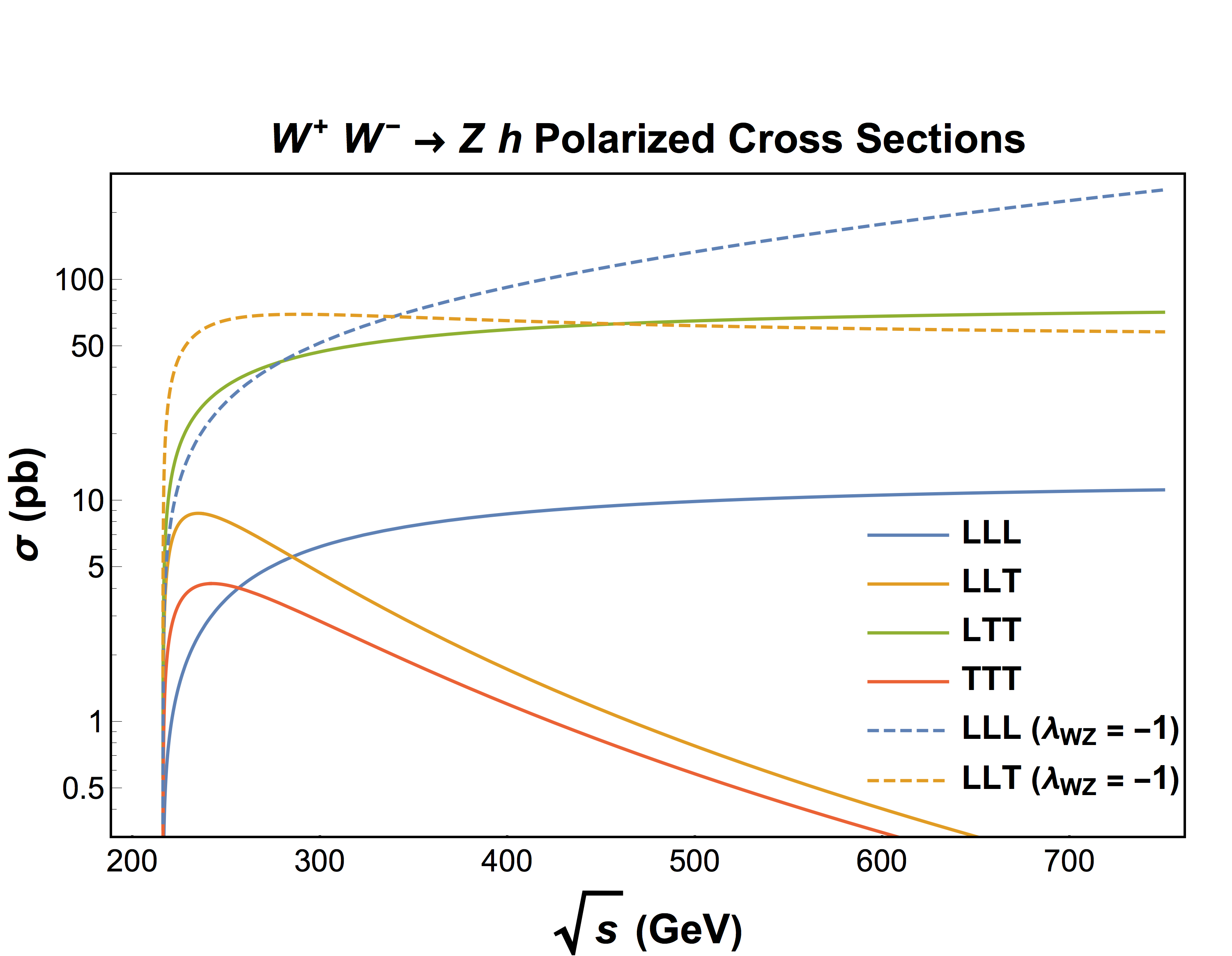}
\includegraphics[width=0.49\textwidth]{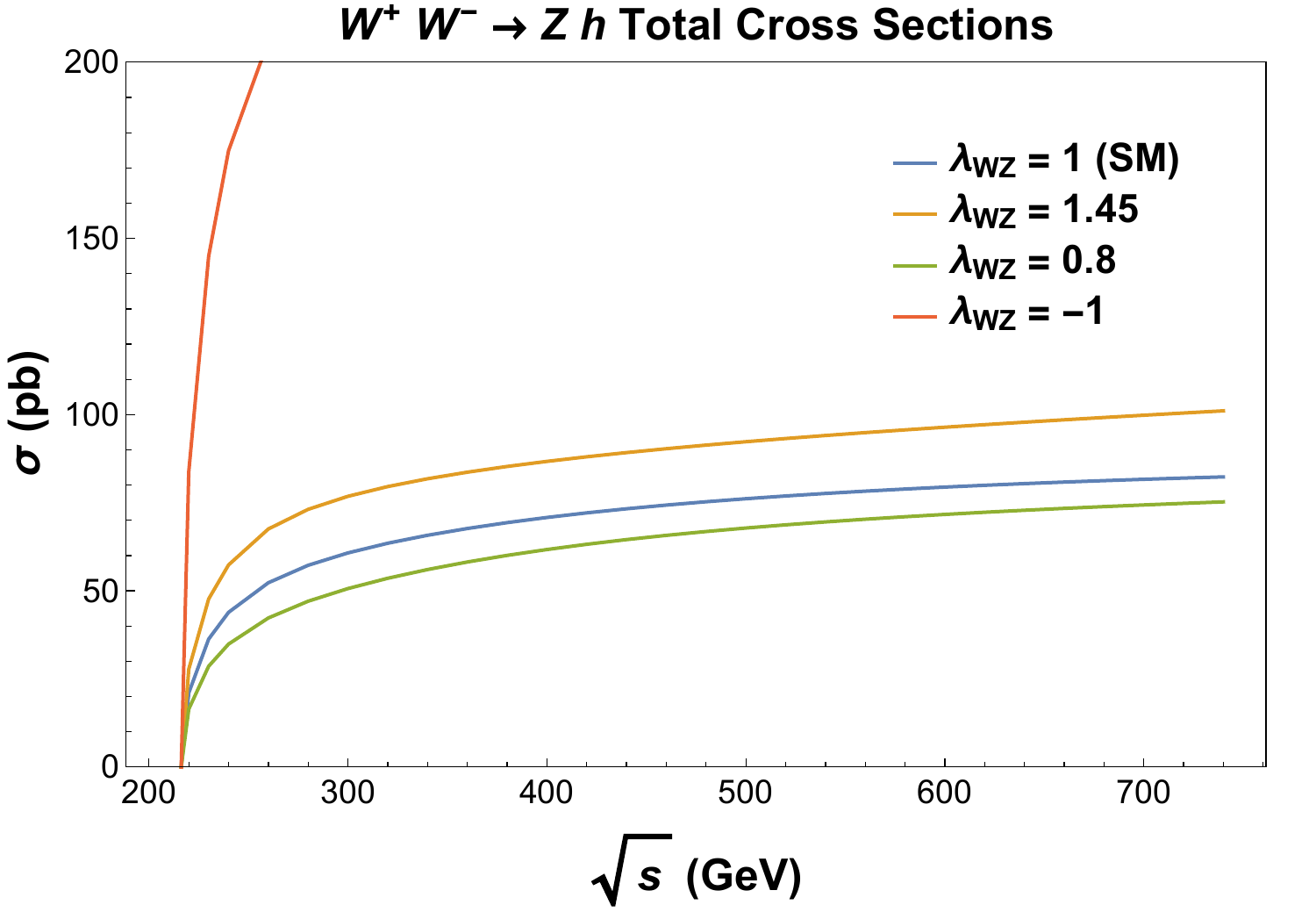}
\caption{\textbf{Left:} Polarized cross sections of the process $W^+W^- \rightarrow Zh$ as a function of center of mass energy $\sqrt{s}$ categorized by the number of transverse (T) or logitudinally (L) polarized vectors in the process. The solid lines are SM results while the dashed lines have $\lambda_{WZ} = -1$. \textbf{Right:} Total cross sections for the same process. The blue line is the SM result, while other lines have different values of $\lambda_{WZ}$. We have taken $\kappa_W = 1$ in all cases. Note that the left plot is on a log-scale, while the right one is linear. }
\label{fig:wwzh-xsec}
\end{figure}

We now look at a full calculation of the cross section in~\autoref{fig:wwzh-xsec}. On the left panel, looking first at the solid lines which are the SM prediction, the cross section is dominated by the process with two transverse and one longitudinal gauge boson (LTT)\footnote{Note that, the energy dependence of the cross section in LTT and LLL configurations does not behave as expected shown in~\autoref{tab:HighEnXsec}. This is mainly due to the fact that in the forward and backward region ($\cos\theta=\pm1$), the high energy expansion is different. When integrating over $\cos\theta$, we will obtain different dependence on the scattering energy. However, the overall cancellation between $s$ and $t/u$ contributions is not ruined.}. Of these, the largest are the processes where one of the initial states is longitudinal, and the two transversely polarized gauge bosons have opposite chirality. If we now turn to the dashed lines which have $\lambda_{WZ}=-1$, we see the dramatic growth with energy of the process with all longitudinal gauge bosons, confirming the analysis of~\autoref{eq:matLLL}. We can also see that the process with two longitudinal gauge bosons is significantly enhanced.

On the right panel of~\autoref{fig:wwzh-xsec} we study how the total cross section as a function of energy varies with $\lambda_{WZ}$, and we can again see that for  $\lambda_{WZ} \simeq -1$, the cross section is much larger than the SM prediction for all energies. Even for moderate modifications of $\lambda_{WZ}$ there can be significant changes to the cross section. Analyzing the isospin related processes $W^\pm Z \rightarrow W^\pm h$ gives analogous results. Therefore we see that because of the two processes that grow with energy, and the cancellation that occurs only at the SM value of $\lambda_{WZ}$, measurement of this process is a very sensitive probe of this coupling ratio. Unfortunately, an electroweak gauge boson collider is not feasible, so directly measuring these processes is not possible. In what follows we turn to the more realistic case of measuring vector boson scattering as a subprocess in a lepton collider.

\section{VH production through VBF}
\label{sec:VBF}

\begin{figure}[!hbt]
\centering
\includegraphics[width=0.99\textwidth]{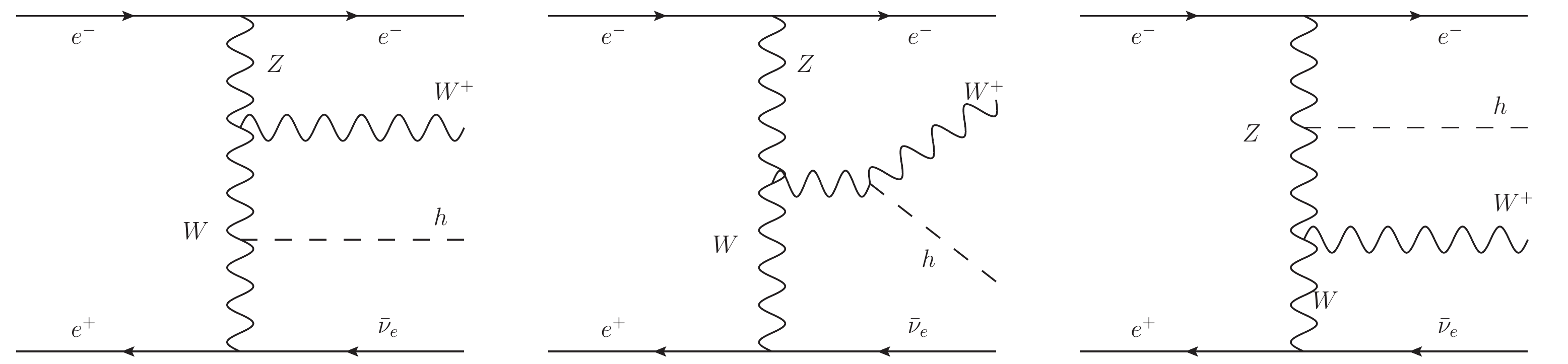}
\hspace{0mm}
\put(-380,-20){\textbf{(a)}}\put(-230,-20){\textbf{(b)}}\put(-80,-20){\textbf{(c)}}\put(-230,-30){\textbf{}}\\
\includegraphics[width=0.99\textwidth]{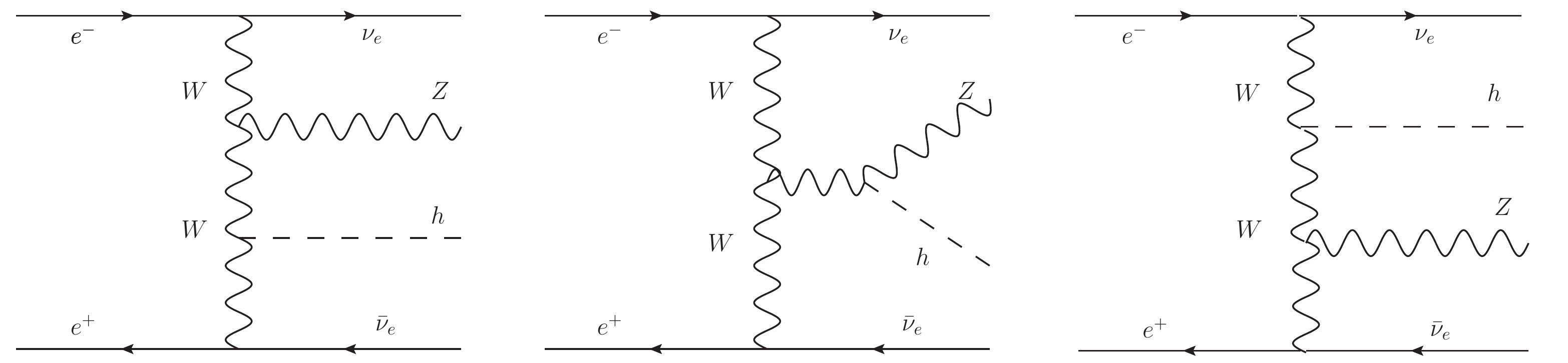}
\hspace{0mm}
\put(-380,-20){\textbf{(d)}}\put(-230,-20){\textbf{(e)}}\put(-80,-20){\textbf{(f)}}
\caption{The Feynman Diagrams for $Wh$ (a,b,c) and $Zh$ (d,e,f) production through VBF processes at Lepton Collider.}
\label{fig:WhZhFeynDia}
\end{figure}

\begin{figure}[!hbt]
\centering
\includegraphics[width=\textwidth]{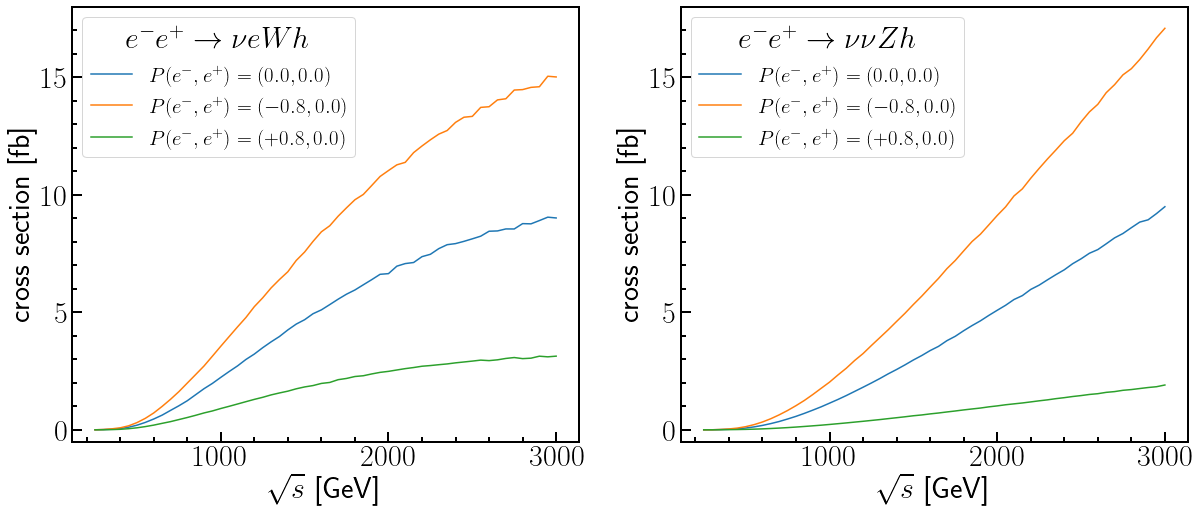}
\caption{The cross section of two processes as a function of $\sqrt{s}$ for there different polarizations. The cross sections are calculated using {\tt MadGraph5\_aMC@NLO}~\cite{Alwall:2014hca} with $p_T^\ell > 10$ GeV and $|\eta^\ell| < 3.5$.}
\label{fig:processesCS}
\end{figure}

The processes we are considering is the $V +$Higgs $(V=W,Z)$ production through VBF at a Lepton Collider\footnote{We use electrons for the initial state, but the analysis is very similar at a muon collider.}:
\begin{align}
e^+\ e^- &\to \nu_e\ \bar{\nu}_e\ Z\ h, \\
e^+\ e^- &\to \nu_e\ e\ W\ h.
\end{align}
The main Feynman diagrams for these two processes are shown in~\autoref{fig:WhZhFeynDia}. \autoref{fig:WhZhFeynDia}(a,b,d,f) contain the $hWW$ vertex, while \autoref{fig:WhZhFeynDia}(c,e) rely on the $hZZ$ vertex. The cross section as a function of collision energy for three different polarizations are also shown in~\autoref{fig:processesCS}. (For the cross section calculation, we require $p_T^\ell>5$ GeV and $|\eta_\ell|<3.5$.) The cross section grows as the collision energy increases. To fully utilize the potential of these two processes, we will consider both the 3 TeV and 1.5 TeV scenarios at CLIC~\cite{Roloff:2018dqu}. We also note that polarization of the electron beam can significantly increase the cross section in the left-handed configuration.  

We denote the contribution of the matrix element square from $hWW$ couplings as $|\mathcal{M}_W|^2$ (e.g. from (a,b) or (d,f) only), the contribution from $hZZ$ couplings as $|\mathcal{M}_Z|^2$ (from (c) or (e) only) and the interference term as $\mathcal{M}_{WZ}^2$. Then the total matrix element square for either $Wh$ or $Zh$ processes can be written as (with the dependence on the $\kappa$ of relevant couplings):
\begin{align}
|\mathcal{M}|^2 = \kappa_W^2|\mathcal{M}_W|^2 + \kappa_W\kappa_Z\mathcal{M}_{WZ}^2 + \kappa_Z^2|\mathcal{M}_Z|^2.
\end{align}
The total cross section is
\begin{align}
\sigma = \kappa_W^2\sigma_W + \kappa_W\kappa_Z\sigma_{WZ} + \kappa_Z^2\sigma_Z.
\label{eq:totalxSec}
\end{align}
Following the polarization configuration presented in~\cite{Roloff:2018dqu}, the individual contributions are listed in~\autoref{tab:CSIndividual} for three different $\sqrt{s}$. We find that, besides the case $\sqrt{s}=350$ GeV where the production rate is small, the interference effect is very large, comparable or even larger than the individual contributions of $\sigma_W$ and $\sigma_Z$. This significant interference effect offers the opportunity to measure the $\kappa_W$, $\kappa_Z$ as well as $\lambda_{WZ}$ precisely. It also demonstrates the dramatic sensitivity to the sign of $\lambda_{WZ}$. In the following section, we conduct a detailed phenomenological study of this channel including backgrounds and realistic experimental cuts. 


\begin{table}[!hbt]
\centering
\begin{tabular}{c|c|rr|rr}
\hline\hline
\multicolumn{2}{c|}{$\sigma$ [fb]} & \multicolumn{2}{c|}{$Wh$} & \multicolumn{2}{c}{$Zh$} \\
\cline{1-2}
$\sqrt{s}$ [GeV] &  & $P(e^-)=-80\%$ & $P(e^-)=80\%$ & $P(e^-)=-80\%$ & $P(e^-)=80\%$\\
\hline
\multirow{3}{*}{350} & $\sigma_Z$ & $6.81\times10^{-3}$ & $2.46\times10^{-3}$ & $1.08\times10^{-2}$ & $2.91\times10^{-3}$ \\
& $\sigma_W$ & $3.85\times10^{-2}$ & $8.27\times10^{-2}$ & $1.49\times10^{-2}$ & $1.65\times10^{-3}$ \\
& $\sigma_{WZ}$ & $-3.94\times10^{-3}$ & $-2.22\times10^{-3}$ & $-1.03\times10^{-2}$ & $-1.16\times10^{-3}$ \\
\hline
\multirow{3}{*}{1500} & $\sigma_Z$ & $8.25\times10^{0}$ & $3.18\times10^{0}$ & $3.85\times10^{0}$ & $4.25\times10^{-1}$ \\
& $\sigma_W$ & $1.22\times10^{1}$ & $4.11\times10^{0}$ & $6.85\times10^{0}$ & $7.66\times10^{-1}$ \\
& $\sigma_{WZ}$ & $-1.28\times10^{1}$ & $-5.46\times10^{0}$ & $-5.38\times10^{0}$ & $5.93\times10^{-1}$ \\
\hline
\multirow{3}{*}{3000} & $\sigma_Z$ & $3.51\times10^{1}$ & $1.34\times10^{1}$ & $1.87\times10^{1}$ & $2.09\times10^{0}$ \\
& $\sigma_W$ & $4.31\times10^{1}$ & $1.50\times10^{1}$ & $2.97\times10^{1}$ & $3.27\times10^{0}$ \\
& $\sigma_{WZ}$ & $-6.32\times10^{1}$ & $-2.52\times10^{1}$ & $-3.13\times10^{1}$ & $-3.45\times10^{0}$ \\
\hline
\hline
\end{tabular}
\caption{The individual contributions to total cross section for $Wh$ and $Zh$ VBF processes at different collision energies and different polarizations. The cross section is obtained from {\tt MadGraph5\_aMC@NLO} with cuts: $p_T^\ell>10$ GeV and $|\eta_\ell|<3.5$. The polarization configuration is following those in~\cite{Roloff:2018dqu}.}
\label{tab:CSIndividual}
\end{table}
%

\section{Phenomenology Study}
\label{sec:pheno}

A measurement of the cross section of the processes studied here can be translated into a measurement of $\lambda_{WZ}$. This process is particularly sensitive to the sign of this parameter because of the tree-level interference, and can thus relatively easily rule out the case with $\lambda_{WZ}<0$ where the destructive interference effect in the SM turns into a constructive effect. Here we perform a realistic study of this cross section measurement. 

\subsection{Total Rate Measurement}
The signal processes we consider are  
\begin{subequations}
\begin{align}
e^-\ e^+ &\to e^{\pm}\ \nu_{e}\ W^\mp\ h, \\
e^-\ e^+ &\to \nu_e\ \bar{\nu}_e\ Z\ h.
\end{align}
\end{subequations}
We consider the final state containing two isolated leptons, two $b$-jets (from Higgs decay) and $\slashed{E}_T$. Thus, the dominant backgrounds\footnote{Another process with the same final states is associated production of a Higgs and two vector bosons, $e^- e^+ \to VVh$. This process also exhibits tree-level interference so it can be thought of as signal rather than background. The cross section is small and the topology is quite different from the signal, so it has a negligible contribution to this analysis after the cuts. } come from
\begin{subequations}
\begin{align}
&e^- e^+ \to t \bar{t} \to b \bar{b} \ell^-\ell^+\nu_\ell \bar{\nu}_\ell,\\
&e^- e^+ \to e^{\pm} \nu_{e} W^\pm Z \to e^\pm \nu_e \ell^\mp \nu_\ell b\bar{b},\\
&e^- e^+ \to \nu_e \bar{\nu}_e Z Z \to \nu_e \bar{\nu}_e \ell^-\ell^+ b\bar{b}, \\
&e^- e^+ \to Z h, Z \to \ell^- \ell^+, h \to b\bar{b},\\
&e^- e^+ \to Z W^+ W^-, Z \to b\bar{b}, W^+\to\ell^+\nu_\ell, W^-\to\ell^-\bar{\nu}_\ell,\\
&e^- e^+ \to Z Z Z, Z\to b\bar{b}, Z\to \ell^-\ell^+, Z\to \nu_\ell \bar{\nu}_\ell.
\end{align}
\end{subequations}


The events are generated using {\tt MadGraph5\_aMC@NLO}~\cite{Alwall:2014hca} with {\tt PYTHIA8}~\cite{Sjostrand:2007gs}  used for showering and hadronization. The detector effects are simulated with {\tt Delphes}~\cite{deFavereau:2013fsa} using the CLIC card~\cite{Leogrande:2019qbe}. In order to improve the sensitivity, we simulate both 3 TeV and 1.5 TeV events with $P(e^-)=-0.8$ for the electron beam which are two scenarios for CLIC with 4000 and 2000 fb$^{-1}$ luminosity respectively~\cite{Roloff:2018dqu}.

\begin{figure}[!tbp]
    \centering
    \includegraphics[width=0.49\textwidth]{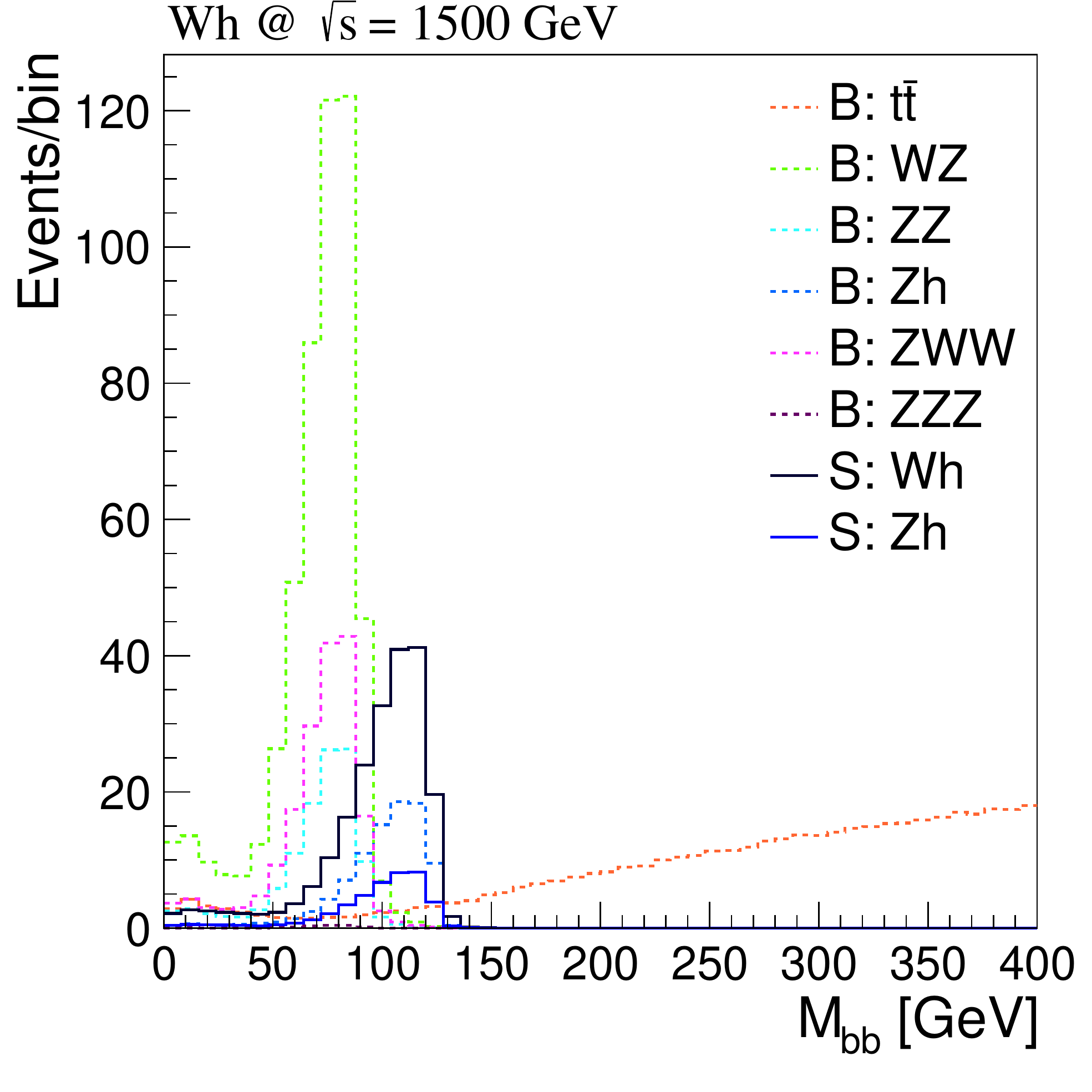}
    \includegraphics[width=0.49\textwidth]{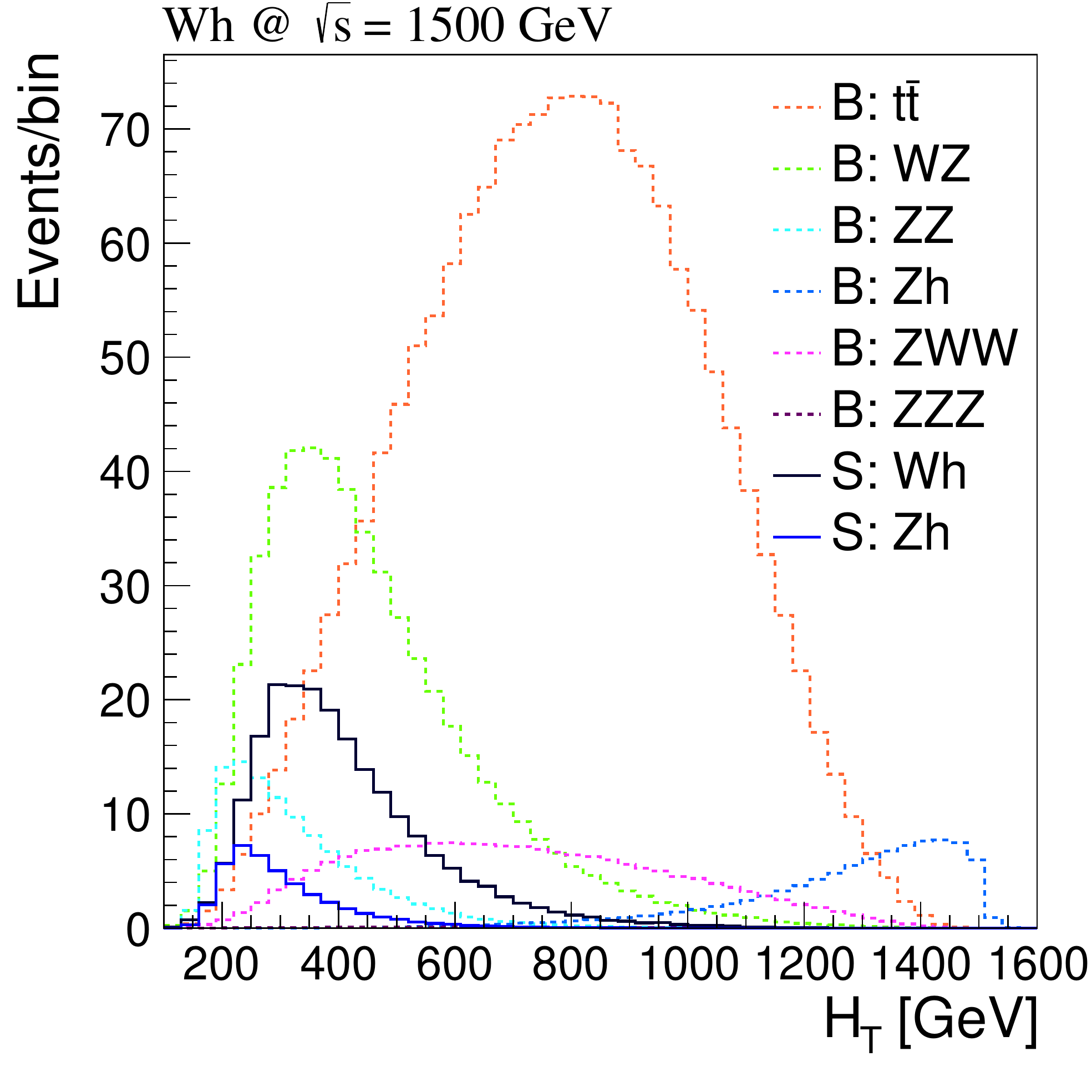}
    \includegraphics[width=0.49\textwidth]{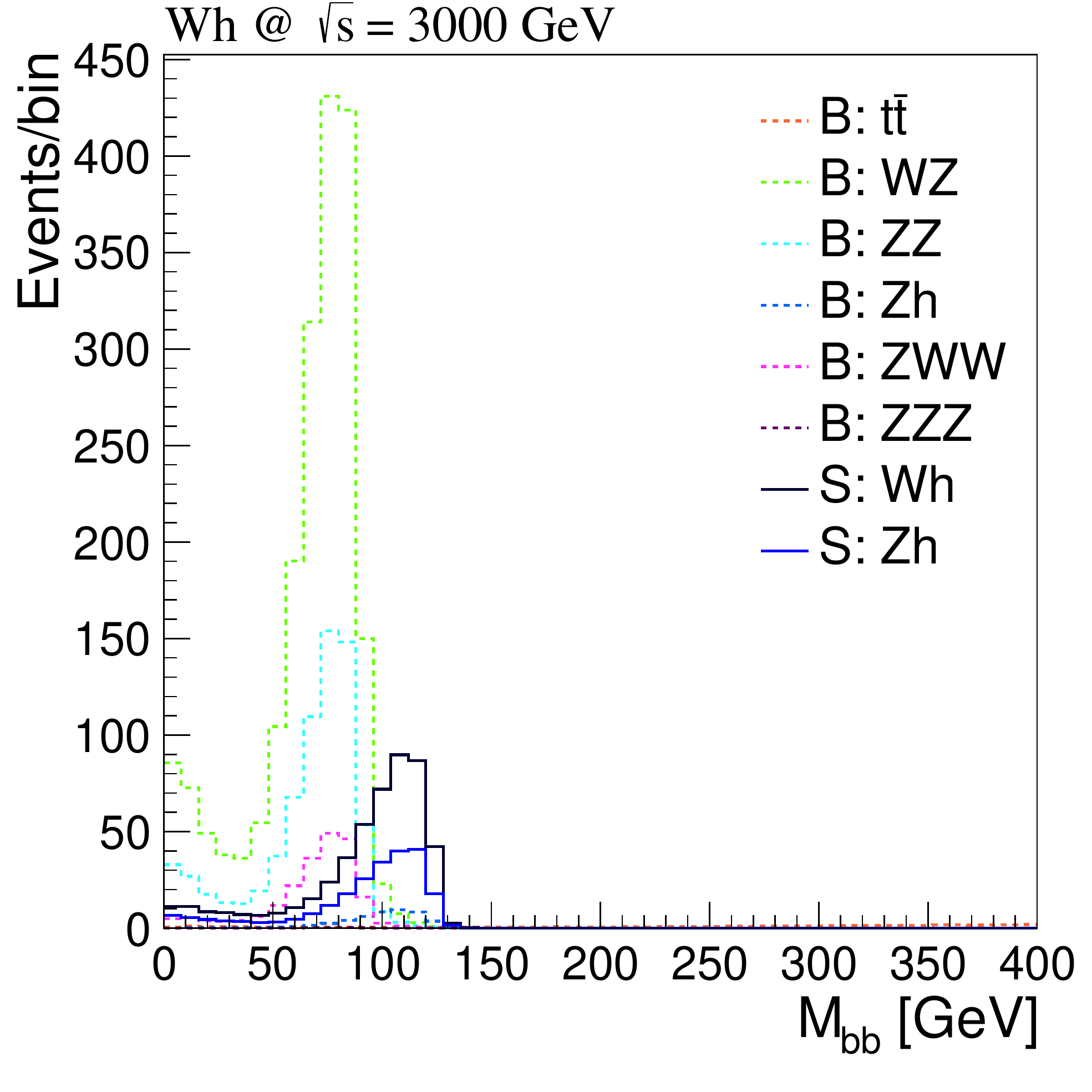}
    \includegraphics[width=0.49\textwidth]{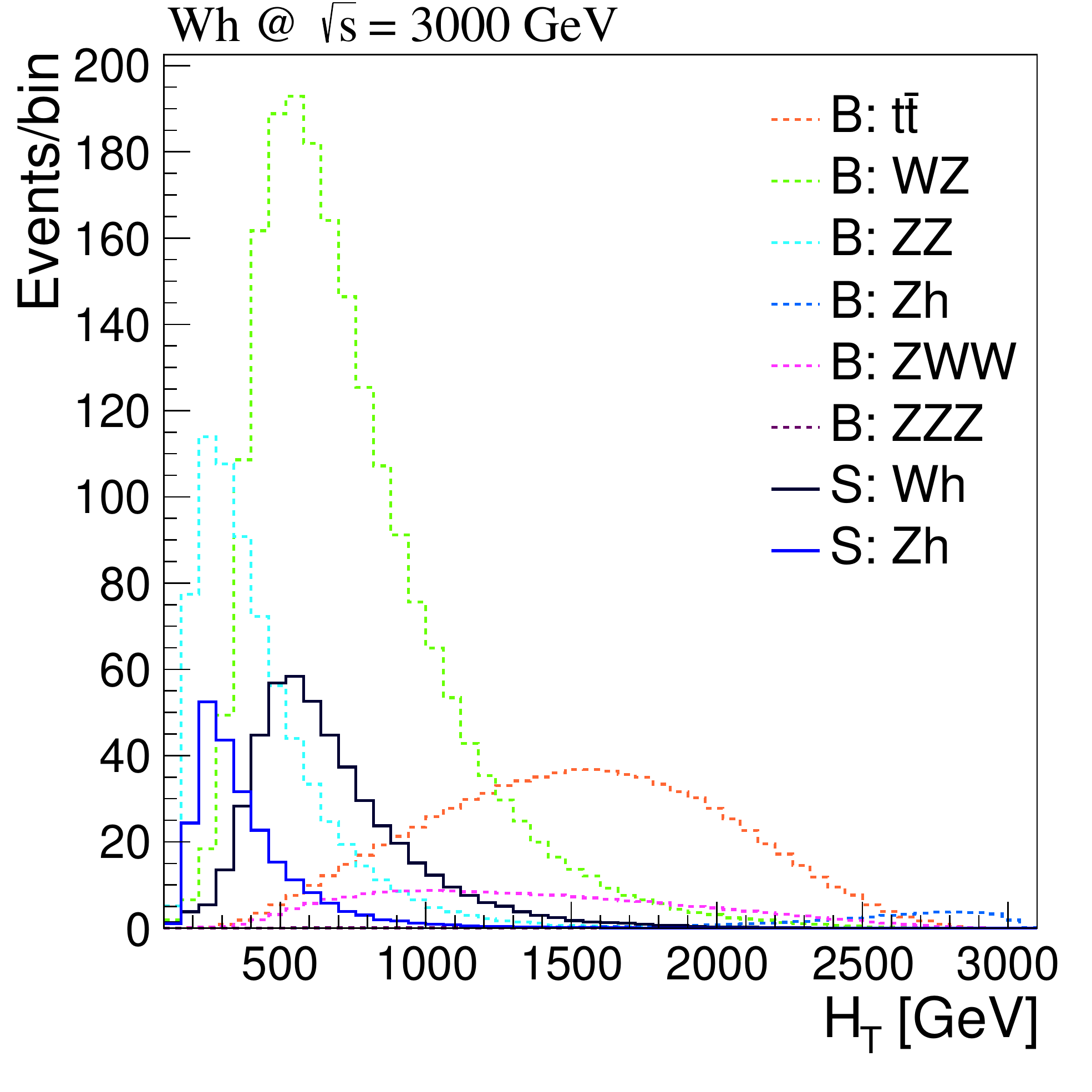}
    \caption{The distribution of $m_{bb}$ and $H_T$ in $Wh$ channel at $\sqrt{s} = 1500$ GeV (Upper panels) and $\sqrt{s} = 3000$ GeV (Lower panels) with only the basic cuts listed in~\autoref{tab:cuts_wh_zh} applied. Note that for $\sqrt{s} = 3000$ GeV the bulk of the $m_{bb}$ distribution for the $t\bar{t}$ process is beyond the horizontal range of the plot.}
    \label{fig:distributions}
\end{figure}

The analysis is separated into two channels aiming on $Wh$ and $Zh$ production respectively. Based on the distributions and the event topology of the signal and background shown in~\autoref{fig:distributions}, the following cuts are applied on the events which are also listed in~\autoref{tab:cuts_wh_zh}:
\begin{itemize}
\item $p_T^\ell > 20$ GeV.
\item $p_T^j > 20$ GeV.
\item Two isolated leptons:
    \begin{itemize}
    \item For $Wh$ channel, at least one electron which directly comes from the beam remnants.
    \item For $Zh$ channel, one pair of opposite-sign same-flavor (OSSF) leptons is required.
    \end{itemize}
\item Two jets tagged as $b$-jet.
\item The invariant mass of the two $b$-jet: $95\ {\rm GeV}\leq m_{bb} \leq 130\ {\rm GeV}$.
\item The invariant mass of the two isolated leptons:
    \begin{itemize}
        \item For $Wh$ channel, $m_{\ell\ell} < 80\ {\rm GeV}$ or $m_{\ell\ell} > 98\ {\rm GeV}$.
        \item For $Zh$ channel, $75\ {\rm GeV} < m_{\ell\ell} < 100\ {\rm GeV}$.
    \end{itemize}
\item The scalar sum of transverse momenta of all reconstructed object:
    \begin{itemize}
        \item For $Wh$ channel:
        \begin{align*}
            \begin{cases}
                H_T \leq 2500\ {\rm GeV} & \sqrt{s} = 3000\ {\rm GeV}, \\
                H_T \leq 1100\ {\rm GeV} & \sqrt{s} = 1500\ {\rm GeV}.
            \end{cases}
        \end{align*}
        \item For $Zh$ channel:
            \begin{align*}
                \begin{cases}
                    H_T \leq 1500\ {\rm GeV} & \sqrt{s} = 3000\ {\rm GeV},\\
                    H_T \leq 700\ {\rm GeV} & \sqrt{s} = 1500\ {\rm GeV}.
                \end{cases}
            \end{align*}
    \end{itemize}
\end{itemize} 

\begin{table}[!tbp]
\centering
\resizebox{\textwidth}{!}{%
\begin{tabular}{|c|c|c|}
\hline
Cuts & $Wh$-Cuts & $Zh$-Cuts \\ \hline
\multirow{3}{*}{Basic Cuts} & \multicolumn{2}{c|}{$p_T^\ell > 20$ GeV, $N_\ell = 2$} \\
    & \multicolumn{2}{c|}{$p_T^j>20$ GeV, $N_b=2$} \\ \cline{2-3}
    & $N_e\geq1$ & 1 OSSF Pair \\ \hline
$m_{bb}$ & \multicolumn{2}{c|}{$95\ {\rm GeV}\leq m_{bb}\leq130\ {\rm GeV}$} \\ \hline
$m_{\ell\ell}$ & $m_{\ell\ell}\leq80\ {\rm GeV}$ or $m_{\ell\ell}\geq98\ {\rm GeV}$ & $75\ {\rm GeV}\leq m_{\ell\ell}\leq 100\ {\rm GeV}$ \\ \hline
$H_T$ & $\begin{cases}H_T\leq 2500\ {\rm GeV} & \sqrt{s} = 3000\ {\rm GeV}\\ H_T\leq 1100\ {\rm GeV}& \sqrt{s} = 1500\ {\rm GeV}\end{cases}$ & $\begin{cases}H_T\leq 1500\ {\rm GeV} & \sqrt{s} = 3000\ {\rm GeV}\\ H_T\leq 700\ {\rm GeV}& \sqrt{s} = 1500\ {\rm GeV}\end{cases}$ \\ \hline
\end{tabular}%
}
\caption{The Cuts used for $Wh$ channel and $Zh$ channel.}
\label{tab:cuts_wh_zh}
\end{table}

The cross sections for all processes after cuts are listed in~\autoref{tab:allcs} for each channel, where in the last row, we also list the expected precision on the measurement of signal cross section with each channel assuming $\mathcal{L} = 4000\,{\rm fb}^{-1}$ for $\sqrt{s} = 3000\,{\rm GeV}$ and $\mathcal{L} = 2000\,{\rm fb}^{-1}$ for $\sqrt{s} = 1500\,{\rm GeV}$. 

\begin{table}[!tbp]
    \centering
    \resizebox{\textwidth}{!}{%
    \begin{tabular}{cccccccc}
    \hline\hline
    \multicolumn{2}{c}{\multirow{2}{*}{$\sigma$ (fb)}} & \multicolumn{3}{c}{$\sqrt{s} = 3.0$ TeV, $\mathcal{L}=4$ ab$^{-1}$} & \multicolumn{3}{c}{$\sqrt{s} = 1.5$ TeV $\mathcal{L}=2$ ab$^{-1}$} \\
    \multicolumn{2}{c}{} & Before Cuts & $Wh$-Cuts & $Zh$-Cuts & Before Cuts & $Wh$-Cuts & $Zh$-Cuts \\ \hline\hline
    \multirow{2}{*}{Signal} & $Wh$(VBF) & $1.97\times10^0$ & $7.26\times10^{-2}$ & $1.36\times10^{-3}$ & $9.62\times10^{-1}$ & $6.54\times10^{-2}$ & $2.37\times10^{-3}$ \\
        & $Zh$(VBF) & $6.47\times10^{-1}$ & $3.49\times10^{-3}$ & $7.21\times10^{-2}$ & $2.03\times10^{-1}$ & $1.30\times10^{-3}$ & $2.87\times10^{-2}$ \\ \hline
    \multirow{6}{*}{BG} & $tt$ & $1.17\times10^0$ & $5.83\times10^{-4}$ & $6.10\times10^{-6}$ & $4.65\times10^0$ & $5.64\times10^{-3}$ & $8.05\times10^{-5}$ \\
        & $WZ$(VBF) & $4.47\times10^0$ & $9.97\times10^{-3}$ & $2.16\times10^{-4}$ & $1.84\times10^0$ & $5.86\times10^{-3}$ & $1.96\times10^{-4}$ \\
        & $ZZ$(VBF) & $1.92\times10^0$ & $4.21\times10^{-4}$ & $8.07\times10^{-3}$ & $5.92\times10^{-1}$ & $1.48\times10^{-4}$ & $2.88\times10^{-3}$ \\
        & $Zh$ & $5.88\times10^{-2}$ & $1.83\times10^{-4}$ & $4.15\times10^{-4}$ & $2.39\times10^{-1}$ & $4.10\times10^{-4}$ & $1.12\times10^{-3}$ \\
        & $ZWW$ & $4.01\times10^{-1}$ & $1.14\times10^{-3}$ & $4.97\times10^{-6}$ & $6.36\times10^{-1}$ & $2.02\times10^{-3}$ & $1.72\times10^{-5}$ \\
        & $ZZZ$ & $5.06\times10^{-3}$ & $6.04\times10^{-7}$ & $1.12\times10^{-5}$ & $9.79\times10^{-3}$ & $1.74\times10^{-6}$ & $2.34\times10^{-5}$ \\ 
        & \textbf{Sum} & $8.02\times10^{0}$ & $1.23\times10^{-2}$ & $8.72\times10^{-3}$ & $7.97\times10^{0}$ & $1.41\times10^{-2}$ & $4.32\times10^{-3}$ \\
    \hline    
        & & Precision (\%) & 6.18 &  6.17 & Precision (\%) &  9.53 &   13.5 \\
        \hline\hline
    \end{tabular}%
    }
    \caption{The cross sections of all signal and background (BG) processes (with final states $b\bar{b}\ell^+\ell^-$) at $\sqrt{s} = 1500,\ 3000$ GeV for $P(e^-) = -0.8$. Note that for the VBF processes, $p_T^\ell > 10$ GeV and $|\eta^\ell|<3.5$ are imposed at the generation level for the forward/backward charged lepton. We also quote the the precision on the measurement of signal cross section that can be extracted with the given luminosity.}
    \label{tab:allcs}
    \end{table}

The numbers listed in~\autoref{tab:allcs} are for $\kappa_W = 1$ and $\kappa_Z = 1$. By assuming that the selection efficiency will not change significantly for different values,\footnote{This is a reasonable assumption, as we didn't use any selection cut that has direct dependence on the values of $\kappa_W$ and $\kappa_Z$.} we can obtain the events at any other values of $\kappa_W$ and $\kappa_Z$ by
\begin{align}
    \mathcal{N}_S(\kappa_W,\kappa_Z) = \mathcal{L}\times\sigma(\kappa_W,\kappa_Z) = \mathcal{L}\times\frac{\sigma^{\rm obs}}{\sigma_0(\kappa_W=1,\kappa_Z=1)}\times\sigma_0(\kappa_W,\kappa_Z),
\end{align}
where $\sigma_0(\kappa_W,\kappa_Z)$ for $Wh$ and $Zh$ at $\sqrt{s} = 1500$ and $3000$ GeV can be constructed from the data listed in~\autoref{tab:CSIndividual} as: $\sigma_0(\kappa_W,\kappa_Z) = \kappa_W^2\sigma_{W} + \kappa_W\kappa_Z\sigma_{WZ} + \kappa_Z^2\sigma_{Z}$. Then, assuming Poisson distribution for observed events, the negative log-likelihood (NLL) function is
\begin{align}
    &\Delta{\rm NLL}(\kappa_W,\kappa_Z) = {\rm NLL}(\kappa_W,\kappa_Z) - {\rm NLL}_{\rm min}\nonumber \\
    = & \sum_i \left(\left(\mathcal{N}_B^i + \mathcal{N}_S^i(1,1)\right)\log\left[\frac{\mathcal{N}_B^i + \mathcal{N}_S^i(1,1)}{\mathcal{N}_B^i + \mathcal{N}_S^i(\kappa_W,\kappa_Z)}\right] + \mathcal{N}_S^i(\kappa_W,\kappa_Z) - \mathcal{N}_S^i(1,1)\right),
\end{align}
where the summation runs over all the channels we have considered. Then the bound with an $N$-$\sigma$ confidence interval corresponds to $\Delta{\rm NLL}\leq N^2/2$.

\begin{figure}
\centering
\includegraphics[width=0.31\textwidth]{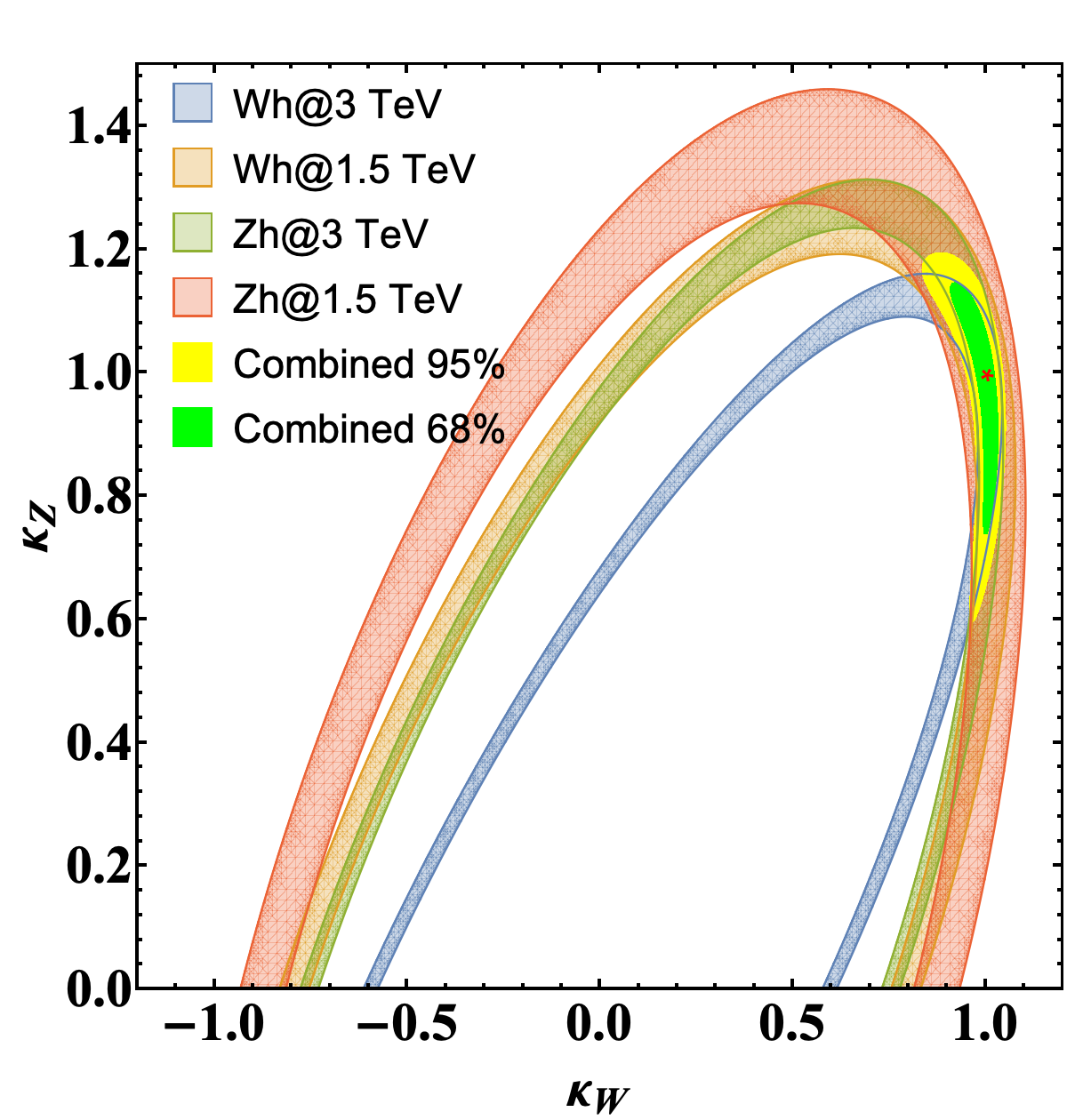}
\includegraphics[width=0.32\textwidth]{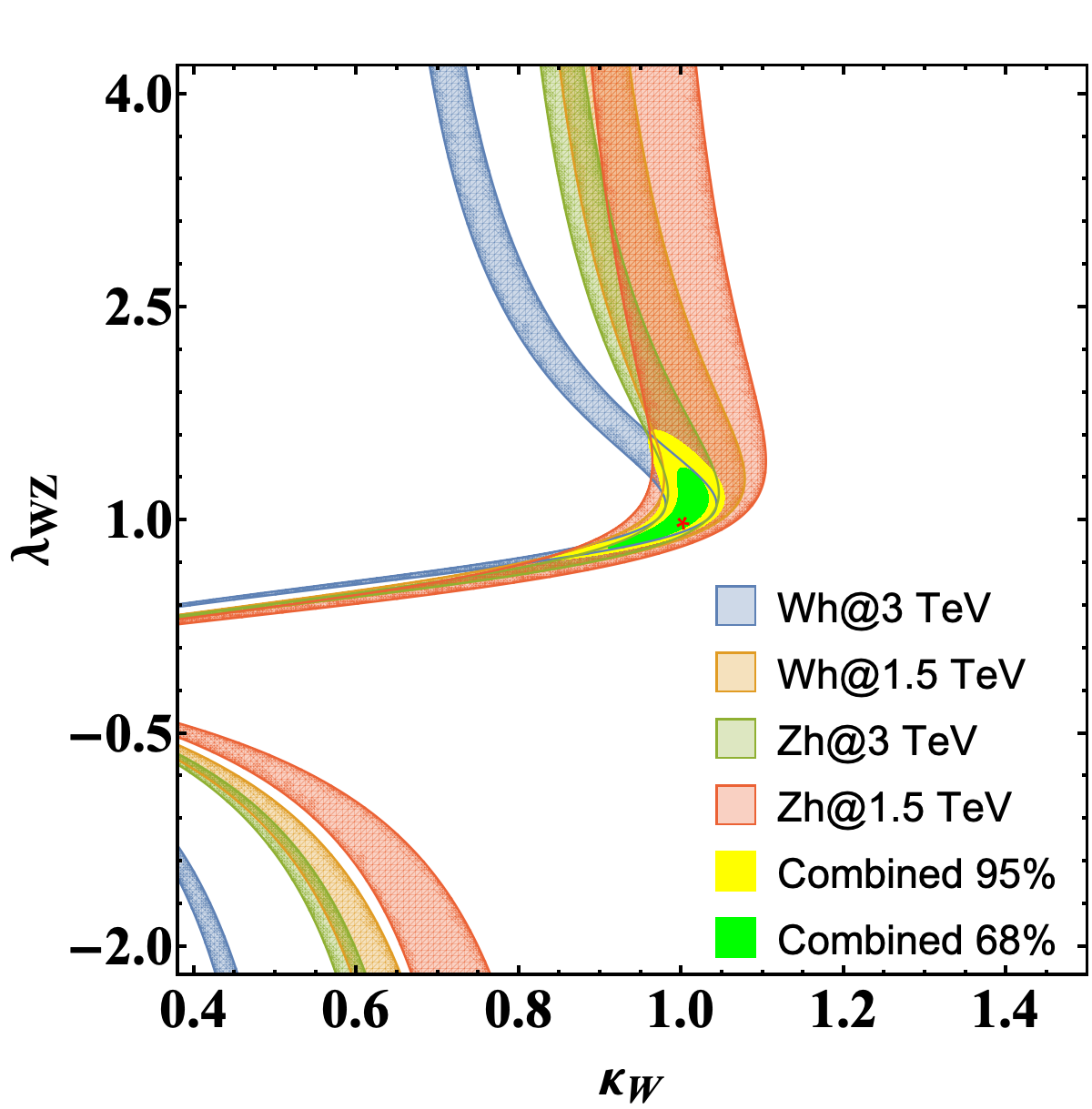}
\includegraphics[width=0.32\textwidth]{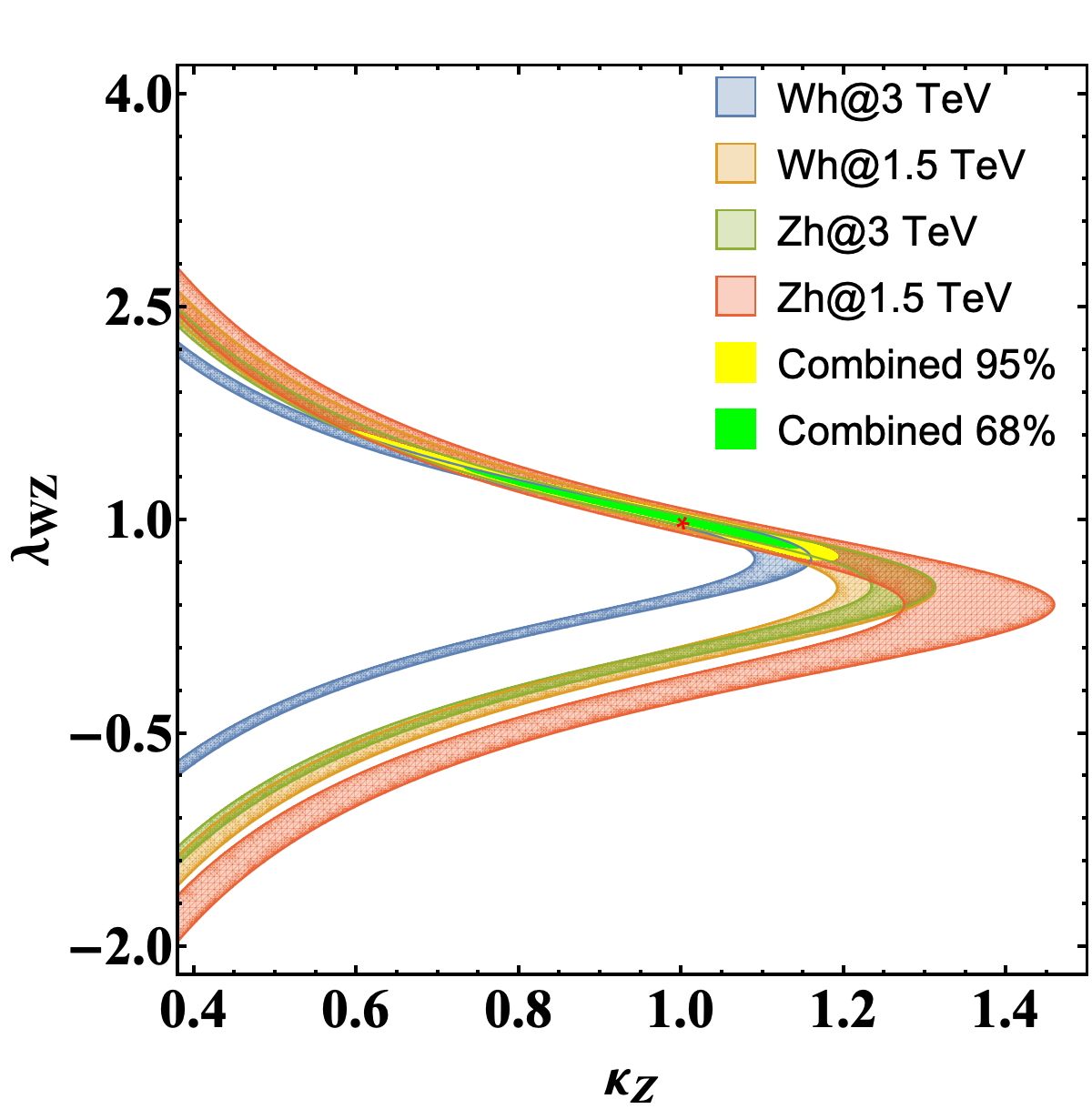}
\caption{The constraints in the $\kappa_W$-$\kappa_Z$, $\kappa_W$-$\lambda_{WZ}$, and $\kappa_Z$-$\lambda_{WZ}$ planes from the total rate measurements. We show the contours from the four different measurements at 68\%, and also show the combined constraints at 68\% C.L. (95\% C.L.) in green (yellow). The SM values are indicated as red points.
}
\label{fig:LamWZtotalrate}
\end{figure}

Combining all these measurements, we can get the 68\% ($\sim1$-$\sigma$) and 95\% ($\sim2$-$\sigma$) confidence level (C.L.) region.
The results are shown in~\autoref{fig:LamWZtotalrate} in $\kappa_W$-$\kappa_Z$ (left panel), $\kappa_W$-$\lambda_{WZ}$ (middle panel) and $\kappa_Z$-$\lambda_{WZ}$ (right panel) plane respectively.
The 68\% region for each individual channel is also given in each plot which shows the complementarity among these channels.
Note that although we didn't optimize the cuts listed in~\autoref{tab:cuts_wh_zh}, they give a sensitivity very close to that achieved with a boosted decision tree using TMVA~\cite{Hocker:2007ht}. 
This is because with these cuts or any relatively similar ones, the signal to background ratio is very high. 

Besides the C.L. region around SM point shown in~\autoref{fig:LamWZtotalrate}, we can also estimate the luminosity that is needed to exclude some non-SM benchmark points. For this purpose, we will not combine $\sqrt{s} = $ 1.5 TeV and 3.0 TeV, as each has its own luminosity. The results are shown in~\autoref{tab:lumExclude}, and we see that significantly less data than the the standard proposals is required to exclude these scenarios. In particular, the scenario with $\lambda_{WZ} \simeq -1$, which is very difficult to probe in other processes, can be probed with just a few inverse femtobarns.

\begin{table}
    \centering
    \begin{tabular}{ccc}
        \hline\hline
        Benchmark & $\sqrt{s} = 3.0$ TeV & $\sqrt{s} = 1.5$ TeV\\\hline\hline
        $\kappa_W = \pm1$, $\kappa_Z = \mp1$ & 3.4 fb$^{-1}$ & 14.1 fb$^{-1}$ \\
        $\kappa_W = 1$, $\kappa_Z = 0 $ & 29.3 fb$^{-1}$ & 243.3 fb$^{-1}$ \\
        $\kappa_W = 0$, $\kappa_Z = 1 $ & 62.1 fb$^{-1}$ & 1772.4 fb$^{-1}$ \\\hline\hline
    \end{tabular}
    \caption{The luminosity that is needed to exclude specific benchmark points at 95\% C.L. against the SM case ($\kappa_W=1$ and $\kappa_Z=1$).}
    \label{tab:lumExclude}
\end{table}

\subsection{Differential Distributions Measurement}

\begin{figure}[!tbp]
    \centering
    \includegraphics[width=0.45\textwidth]{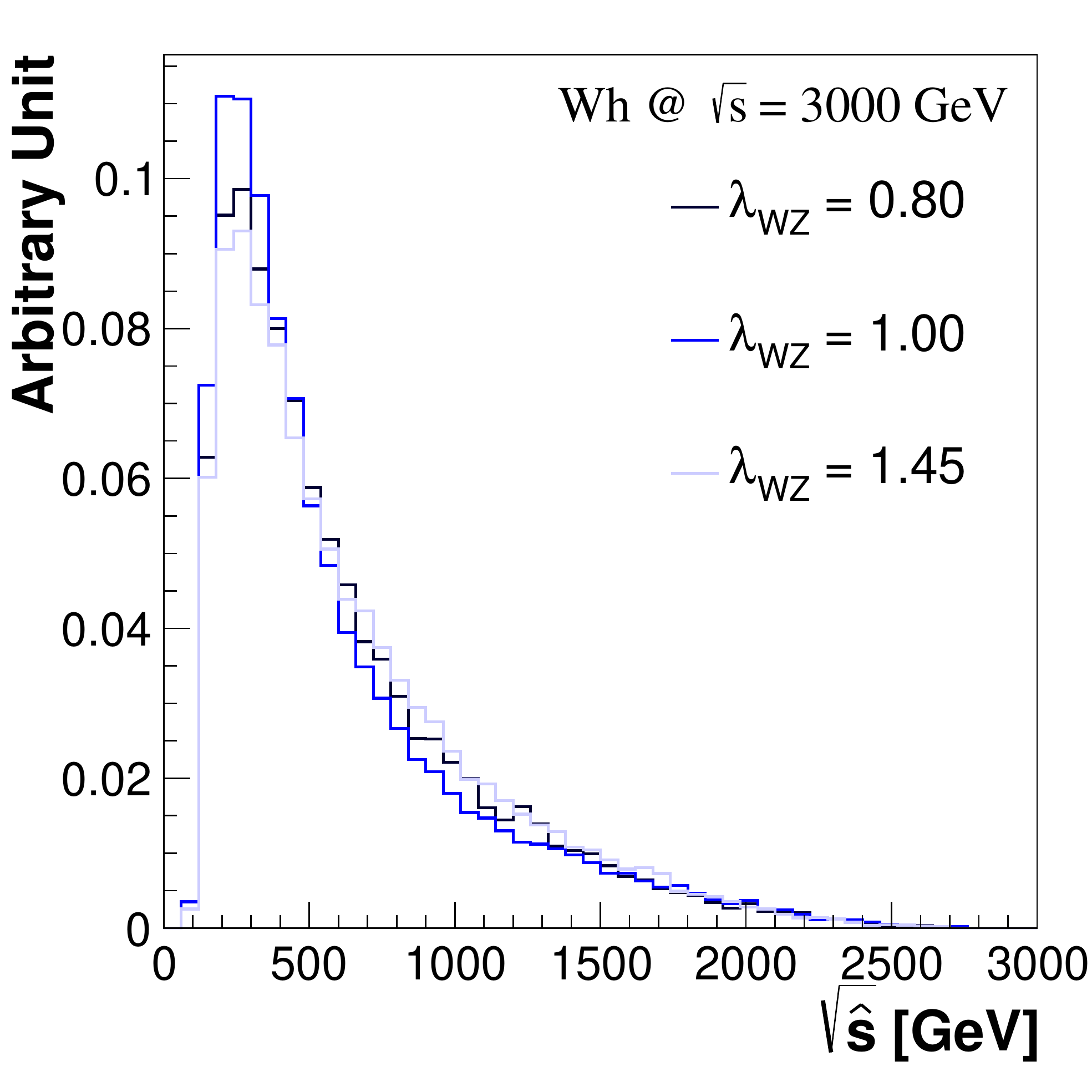}
    \includegraphics[width=0.45\textwidth]{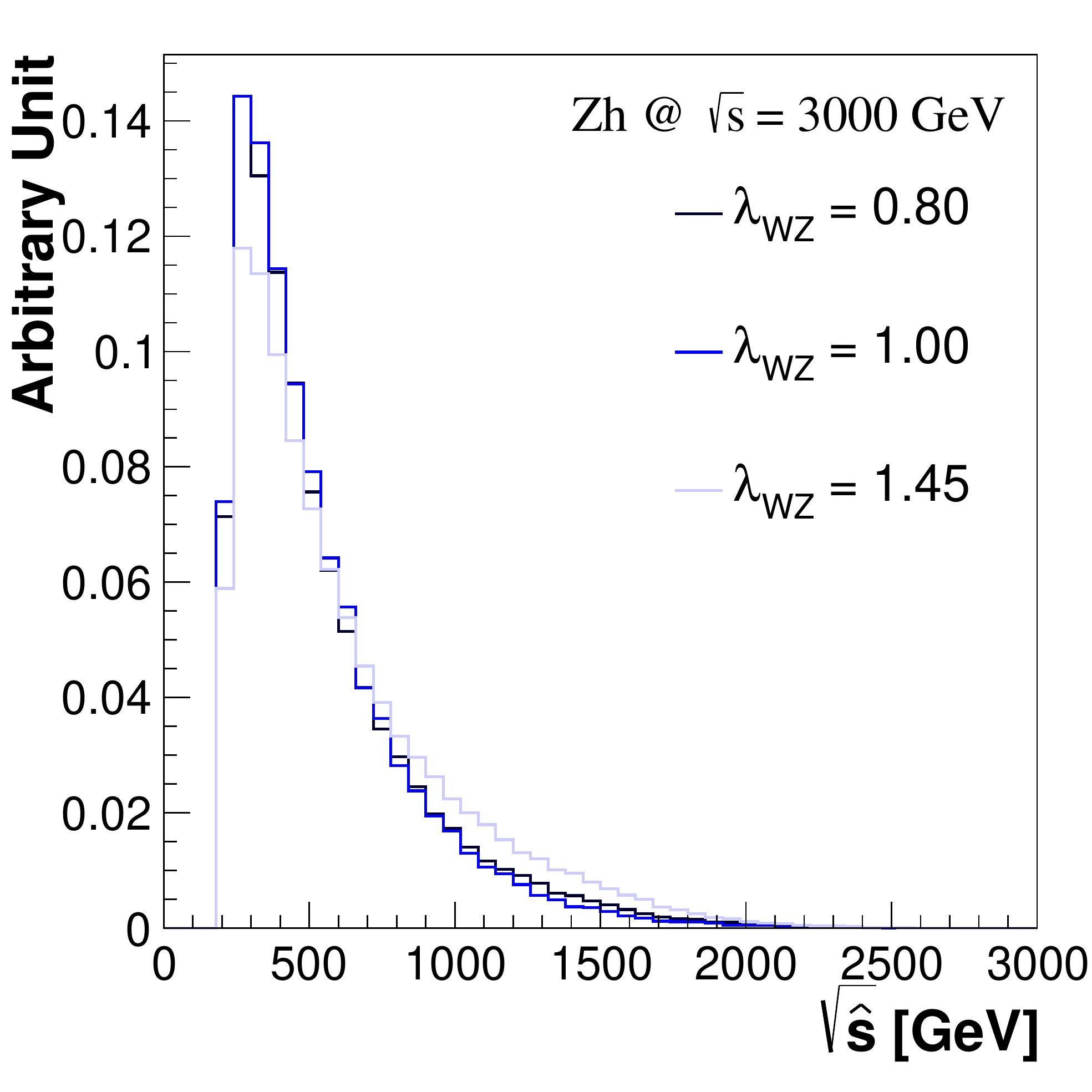}
    \caption{The differential distributions of $\hat{s}$ for different values of $\lambda_{WZ}$ for $Wh$(left) and $Zh$(right) channel at $\sqrt{s} = 3000$ GeV.}
    \label{fig:Shat_Distribution}
\end{figure}

Because of the growth with energy described in~\autoref{sec:2to2}, the dependence of the matrix element on $\hat{s}$ is different for $s$- and $t$/$u$-channels. Therefore, the $s$ and $t$/$u$ channels depend on $\kappa_W$ and $\kappa_Z$ differently. Hence, the differential distribution shape will shift for different value of $\lambda_{WZ}$. This is illustrated in~\autoref{fig:Shat_Distribution} in which we choose 3 different values of $\lambda_{WZ}$ with $\kappa_{W} = 1$ and the distributions are all renormalized to 1. Note that we have generated sufficiently large Monte Carlo samples to suppress the statistical fluctuations. Thus, the difference in the shape of the distributions for different values of $\lambda_{WZ}$ comes from the behaviour of the matrix element. 

\begin{figure}[!tbp]
    \centering
    \includegraphics[width=0.6\textwidth]{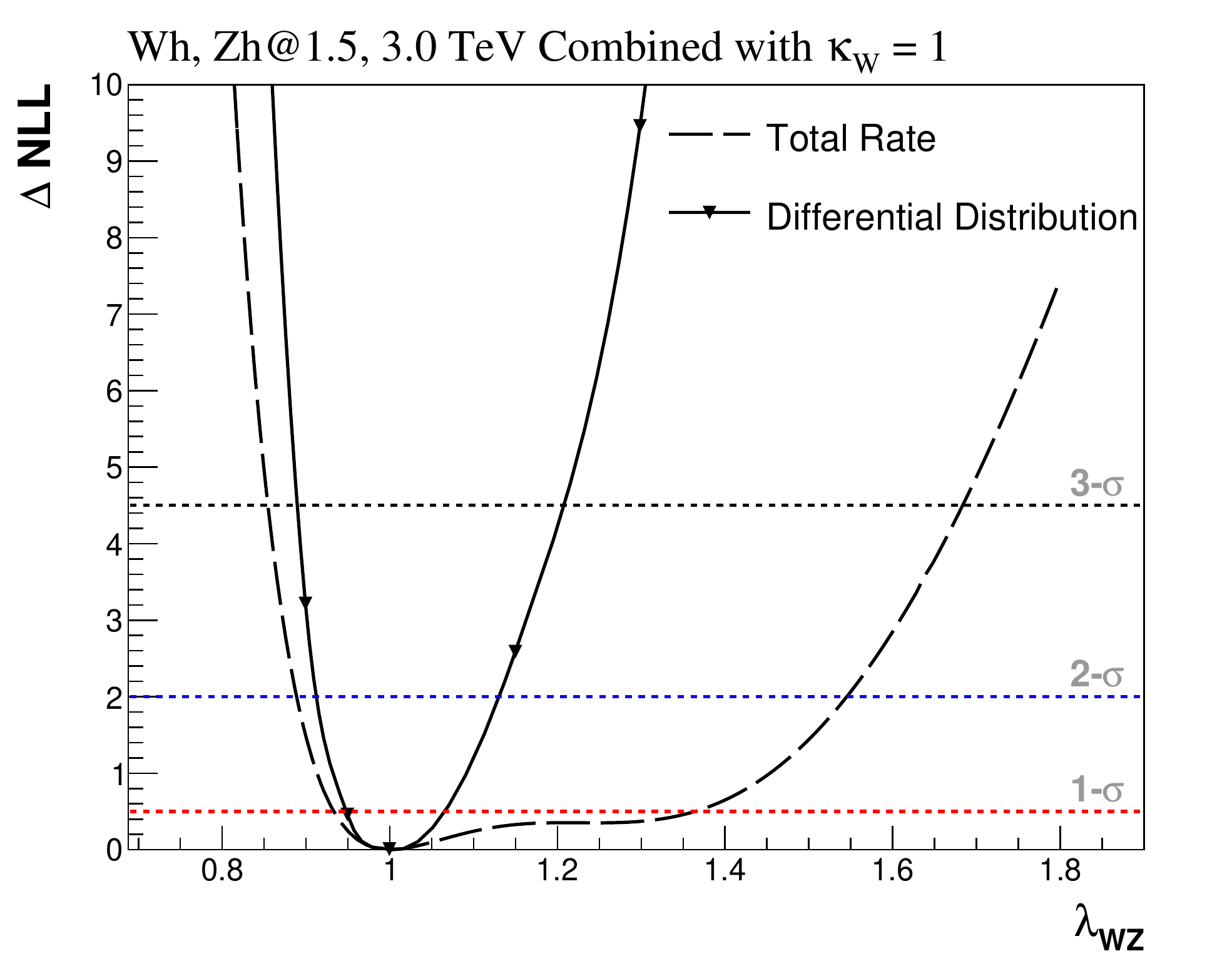}
    \caption{The $\Delta$NLL (see text for definition) for different values of $\lambda_{WZ}$ with $\kappa_W=1$ combining all $Wh$ and $Zh$ channels at both 1.5 and 3.0 TeV CLIC.}
    \label{fig:DNLL}
\end{figure}

In order to estimate the discrimination power, the MC events (including both signal and backgrounds) are used to obtain the ``observed'' ($\kappa_W=1$ and $\lambda_{WZ}=1$) and ``expected'' (otherwise) distributions (shape) of $\sqrt{\hat{s}}$. The distributions as well as the total number of events are utilized to construct the extended likelihood by which we can determine the C.L. region for $\lambda_{WZ}$ using similar method in previous section.  The $\Delta{\rm NLL}$ as a function of $\lambda_{WZ}$ when $\kappa_W=1$ is shown in~\autoref{fig:DNLL}. We find that utilizing the $\sqrt{\hat s}$ distribution can significantly improve the sensitivity.

From~\autoref{fig:DNLL}, we see that the sensitivity of the total measurement is particularly weak for $1 \lesssim \lambda_{WZ} \lesssim1.5$. This is because for fixed $\kappa_W$ as in~\autoref{fig:DNLL}, we have from \autoref{eq:totalxSec}
\begin{equation}
\frac{\partial \sigma}{\partial \lambda_{WZ}} \sim -\frac{\sigma_{WZ}}{\lambda_{WZ}^2}- \frac{2\sigma_{Z}}{\lambda_{WZ}^3}.
\end{equation}
Looking at~\autoref{tab:CSIndividual}, we see that in all four scenarios we are interested in, this derivative vanishes for $\lambda_{WZ}$ between 1 and 1.5, so the cross section in that region is changing very slowly and there is little sensitivity. On the other hand, the differential cross section in $\sqrt{\hat{s}}$ will change with fewer events near threshold and more events at higher energy as we move away from $\lambda_{WZ}=1$, allowing this analysis to break the approximate degeneracy in the total rate measurement.


\section{Summary}
\label{sec:sum}

The nature of electroweak symmetry breaking and whether the Higgs has the precise properties given in the Standard Model is still not fully explored by data. The couplings of the Higgs to heavy gauge bosons are a particularly important probe of the mechanism that gives mass to the $W$ and $Z$. The ratio of those couplings, $\lambda_{WZ}$ (see \autoref{eq:lambda} and \autoref{eq:lag} for a precise definition) has been measured by the LHC to approximately 10\% precision, but as yet we have essentially no information about the sign of $\lambda_{WZ}$. 

Sign measurements are easiest in processes with tree-level interference; these processes will have very large deviations if the relative sign of a coupling is changed but the magnitude is kept constant. In this work we have studied $VV\rightarrow Vh$, $V=W,Z$ using vector boson fusion at a high energy lepton collider. This process exhibits tree-level interference between diagrams proportional to the Higgs coupling to $W$ and that to $Z$. If there is a deviation from the SM prediction of $\lambda_{WZ}=1$, then this process exhibits growth with energy. Therefore, at high energy there will be very large destructive interference between different processes as shown in \autoref{tab:CSIndividual}. 

We have performed a study of this process at a potential future lepton collider with centre of mass energies of 1.5 and 3 TeV. With simple cuts, one can get a signal to background ratio well above one. We have shown that combining the 
$e^+\ e^- \to \nu_e\ \bar{\nu}_e\ Z\ h$ and the 
$e^+\ e^- \to \nu_e\ e\ W\ h$ channels as well as combining measurements at different center of mass energies, one can measure the couplings of the gauge bosons to the Higgs with a reasonable precision. 
As discussed above, this channel is particularly powerful at probing $\lambda_{WZ}\simeq -1$, which can be excluded with a small fraction of the expected data. Finally, we have shown that including the 
$\sqrt{\hat{s}}$ differential distribution can further improve the measurement. 

\begin{acknowledgments}
We would like to thank Cheng-Wei Chiang for very useful discussions and communication. This work is supported in part by the Natural Sciences and Engineering Research Council of Canada (NSERC).
\end{acknowledgments}

\bibliographystyle{bibsty}
\bibliography{references}

\end{document}